
\parindent=10pt
\baselineskip=15pt
\parskip=5pt

\centerline{\bf $ \rho $ PARAMETER IN THE VECTOR CONDENSATE MODEL}
\centerline{\bf OF ELECTROWEAK INTERACTIONS}
\vskip 2truecm

\centerline{G.Cynolter, E.Lendvai and G.P\'ocsik}
\centerline{Institute for Theoretical Physics}
\centerline{E\"otv\"os Lor\'and University, Budapest}

\vskip 1truecm

\centerline{ABSTRACT}

{\hsize=5.25in \
\leftskip=1 in     In the standard model of
electroweak interactions the Higgs doublet is
replaced by a doublet  of vector bosons and the gauge symmetry is broken
dynamically. This generates masses for the gauge bosons and fermions, as
well as it fixes the interactions in the model. The model has a low momentum
scale. In this note we show that the model survives the test of the $ \rho $
parameter, and to each momentum scale $ \rho $ chooses a possible range of
vector boson masses. \par}

\vskip 1truecm

     Although the Higgs mechanism is a very simple realisation of symmetry
breaking, alternative realisations [1] are also important, since the Higgs has
not yet been seen.

     Recently, a model of symmetry breaking has been introduced [2] in such a
way that in the standard model of electroweak interactions the Higgs doublet
was replaced by a Y=1 doublet of vector fields,
$$ B_{ \mu } = \pmatrix{ B_{ \mu }^{(+)} \cr
B_{ \mu }^{(0)} \cr} , \eqno (1) $$
and $ B_{ \mu }^{(0)} $ forms a nonvanishing condensate,
$$ \left \langle B_{ \mu }^{(0)+} B_{ \nu }^{(0)} \right \rangle_0
= g_{ \mu \nu } d , \qquad d \not = 0 , $$
$$ \left \langle B_{ \mu }^{(+)+} B_{ \nu }^{(+)} \right \rangle_0 = 0 .
\eqno (2) $$
This generates mass terms for W and Z, the tree level $ \rho $ parameter
 remains 1 and the photon is massless. Assuming a quartic self-interaction
for $ B_{ \mu } $ , the $ B^{+,0} $ particles become massive
and the ratio of their bare masses is $ ( 4 / 5 )^{1/2} $.

     Fermions can easily be made massive by introducing
$ \overline{ \Psi }_L B_{ \nu } \Psi_R B^{(0) \nu + }$ + h.c.
type interactions. Also the Kobayashi-Maskawa mechanism can be built in.
Low energy charged current phenomenology shows that
$ \sqrt{-6d}=246 $ GeV. The fermion-B coupling is $ 3 (2)^{ -1/2 } m_f G_F $,
a $ G_F^{1 \over 2} $ factor weaker than the coupling to the Higgs. The model
 can be considered as a low energy effective model valid up to about
$ \Lambda \leq 2.6 $ TeV, and $ m_B \geq  $ 43  GeV [2].

   The aim of this paper is to show that oblique radiative corrections [3]
due to B-loops can give arbitrarily small contributions to the
$ \rho $ parameter, provided the $ B^{+,0} $  masses are
suitably chosen. The cutoff $ \Lambda $ remains unrestricted.

     The contribution $ \Delta \rho $ due to B-loops to  $ \rho $ is
$$ \Delta \rho = \alpha T = { \varepsilon }_1   \eqno (3) $$
where T (or $ { \varepsilon }_ 1 $) is one of the three parameters [3]
constrained by precision experiments. The analysis in Ref. [4]
finds for beyond the standard model $ \Delta \rho = -(0.009 \pm
0.25) 10^{-2} $ at $ m_t =130 $ GeV,  $ m_H =m_Z $.

     The parameter T is defined by
$$ \alpha T=  { e^2 \over {s^2 c^2 m^2_Z } }  \Bigl(
\overline{ \Pi}_{WW} (0) -
\overline{ \Pi }_{WW} (0) \Bigr) \eqno (4) $$
with $ s= \sin \theta _W , c= \cos \theta _W $ is calculated in
one-B-loop order.  $
\overline {\Pi }_{ik} $ is expressed by the $ g_{\mu \nu} $
terms of the vacuum polarization contributions $ \overline{ \Pi
}_{ik} $ due to B-loops as $$ \Pi _{ZZ} = {e^2 \over { s^2 c^2 }
} \overline{ \Pi}_{ ZZ} ,\quad
\Pi_{ WW } = {e^2 \over { s^2 } } \overline{ \Pi}_{ WW } \eqno (5)$$
The interactions giving rise to W and Z self energies follow
from the starting gauge invariant Lagrangian of the  $ B^{+,0} $
fields [2]. From these only the trilinear ones are important to
one-loop order,

$$ L\left( B^{0} \right)={ig \over 2c } \partial^{\mu} { B^{(0)
\nu }}^+ \left( Z_{ \mu} B_{\nu}^{(0)} - Z_{\nu} B^{(0)}_{\mu}
\right) + h.c. , $$ $$ L\left( B^+ {B^+}^+ Z \right)=-C \cdot L\!
\left(B^0 \rightarrow B^+ \right),
\quad C=c^2-s^2, \eqno (6) $$
$$\eqalign{ L \left( B^0B^+W \right) &= {ig \over \sqrt{2} } \biggl[
\partial^{\mu} { B^{(+) \nu }}^+ \left( W_{\mu}^+ B_{\nu}^{(0)} -
W_{\nu}^+ B_{\mu}^{(0)} \right) + \cr
&+ \partial^{\mu} { B^{(0) \nu }}^+ \left( W_{\mu}^- B_{\nu}^{(+)} -
W_{\nu}^- B_{\mu}^{(+)} \right) \biggr] +h.c.\ . \cr } $$
In a renormalizable theory T is finite. In the present model,
however, it is a function of the cutoff  $ \Lambda $ which cannot be
removed. After lengthy calculations we get
$$ \eqalign{ -64 \pi^2 \overline{ \Pi}_{ZZ}(0) = &- {5 \over 4} \left(
{1 \over m_0^2} + { C \over m_+^2 } \right) \Lambda^4 +{11 \over 2}
(1+C) \Lambda^2 -{17 \over 2} m_0^2 \ln \Bigl( 1+{\Lambda^2 \over
m_0^2} \Bigr)- \cr
& - {17 \over 2} C m_+^2  \ln \Bigl( 1+{\Lambda^2 \over m_+^2}
\Bigr) + 3(m_0^2+Cm_+^2)-{3 m_0^4 \over \Lambda^2 +m_0^2} -{3 C
m_+^4 \over \Lambda^2 +m_+^2} \cr } \eqno (7) $$ and

$$ \eqalign{ - \left[ {1 \over 256 \pi^2 } \left( {1 \over m_0^2}+
{1 \over m_+^2} \right) \right]^{-1} \overline{\Pi}_{WW}(0) = -5
\Lambda^4 +11(m_0^2+m_+^2) \Lambda^2 - {(m_0^2-m_+^2)^2 \over
m_0^2+m_+^2 }\Lambda^2 &- \cr - {1 \over 4} \left[
17(m_0^2+m_+^2)^2+3(m_0^2-m_+^2)^2 \right] \ln \left(1+{\Lambda^2
\over m_0^2} \right) \left( 1+ {\Lambda^2 \over m_+^2}
\right) &+ \cr
 {1 \over 2} \left[ -{15 \over 2}(m_0^2+m_+^2)(m_0^2-m_+^2) +{1 \over 2}
{(m_0^2-m_+^2)^3 \over (m_0^2+m_+^2) }-3{(m_0^2+m_+^2)^3 \over (m_0^2-m_+^2)}
 \right] \ln {1+{\Lambda^2 \over m_0^2} \over
  1+ {\Lambda^2 \over m_+^2}  } & } \eqno (8) $$
where $ m_+ (m_0) $  is the physical mass of $ B^+(B^0) $.

     From (3), (4), (7), (8) it follows that $ \Delta \rho $
cannot be small for $ \Lambda^2  \gg m_0^2 , m_+^2 $   and there is no
compensation for $ \Lambda^2 \sim m_0^2 \gg m_+^2 $ or
$ \Lambda^2 \sim m_+^2 \gg m_0^2 $ .
There exists, however, a compensation between the vacuum
polarizations if $ \Lambda,\ m_0,\ m_+ $,   are not very
different.  This is seen in Fig.1 showing $ \Delta \rho $  as the function of
$ m_0 $ at various $ \Lambda $'s for $ m_+=0.8 m_0 $ .  Similar curves
can be drawn for not very different $ m_+ /m_0 $  , but at
higher $ \Lambda $ the sensitivity to   $ m_+ /m_0 $   increases
(Fig.2). For instance, if $ \Lambda=2 $ TeV and $ m_+=0.9 m_0
\pm 10 $ \% , then $ m_0 =1557 \pm 50 $ GeV at $ \Delta \rho =0 $.
We have checked numerically the compensation up to $ \Lambda =15 $
TeV. Increasing $ \Lambda \quad m_0 (\Delta \rho =0) $ grows and the
$ m_0 $ interval corresponding to $ \Delta \rho_{exper.} $
shrinks. For instance, for  $ m_+=0.9 m_0 $  and at $
\Lambda=(0.8, 1.4, 2, 4) $ TeV $ m_0 (\Delta \rho =0) =(623,
1090, 1557, 3115 ) $ GeV and from  $ \Delta \rho_{exper.} $
the $ m_0 $  interval is (619- 630, 1088-1094, 1556-1560,
3114-3116) GeV, etc.

     In conclusion, the vector condensate model does not contradict to
the experimental $ \rho $ parameter and even  $ \Delta
\rho_{exper.} =0 $ can be included, but
this is possible only if the B particle is heavy.

 This work is supported in part by OTKA I/3, No. 2190.

\

REFERENCES

1. Proceedings of LHC Workshop, Aachen 1990, CERN 90-10, Eds. G.Jarlskog and
    D.Rein, Vol. 2, p.757. M.Lindner, Int. Journal of Mod. Phys. A8 (1993)
2167.
    R.Casalbuoni, S.De Curtis and M.Grazzini, Phys. Lett. B 317 (1993) 151.

2. G.P\'ocsik, E.Lendvai and G.Cynolter, Acta Phys. Polonica B 24 (1993) 1495.

3. D.Kennedy and B.W.Lynn, Nucl. Phys. B 322 (1989) 1. M.E.Peskin and
    T.Takeuchi, Phys. Rev. Lett. 65 (1990) 964. G.Altarelli and R.Barbieri,
Phys.
    Lett. B 253 (1990) 161. M.E.Peskin and T.Takeuchi, Phys. Rev. D 46 (1992)
381.

4. J.Ellis, G.L.Fogli and E.Lisi, Phys. Lett. B 292 (1992) 427.

\vskip 2truecm

FIGURE CAPTIONS

Fig. 1    $ \Delta \rho $   vs. $ m_0 $ at various $ \Lambda $'s for $
m_+=0.8 m_0 $  .

Fig. 2.    $ \Delta \rho $   vs. $ m_0 $ at various $ \Lambda $'s for $
m_+=m_0 $  .

\bye